# Effect of the acrylic acid content on the permeability and water uptake of latex films.

Yuri Reyes-Mercado[1], Francisco J. Rodríguez-Gómez[1] and Yurko Duda[2]

[1] Facultad de Química, UNAM, 04510 México, D.F. México.

[2] Programa de Ingeniería Molecular, Instituto Mexicano del Petróleo, 07730 México, D.F. México

**Abstract**
Acrylic acid (AA) is a monomer commonly employed in emulsion polymerization to provide electrostatic colloidal stability and improve specific film performance. The addition of AA not only modifies the kinetics of the polymerization, but also it takes part in the interaction between colloidal particles, which has a strong influence on their packing and consequent latex film properties. In this contribution a theoretical modeling of the latex film formation is presented and compared to experimental results: water vapor permeability and latex film capacitance are studied as a function of AA content. It has been shown that water uptake is mainly affected by film morphology which in turn is defined by intercolloidal interaction and drying rate.

**1. Introduction**
Latex is an aqueous dispersion of polymeric particles that is employed to produce coatings, paints, adhesives, etc. Most of these applications involve the formation of a film from the polymeric dispersion. It has been pointed out that the use of latex will grow beyond the current uses because their low or null environmental impact, since these systems employ water as dispersive medium [1].

Lattices are commonly obtained by emulsion polymerization reactions. In this polymerization technique a monomer or a mixture of them are emulsified in water by means of a surfactant. In order to ensure an adequate reaction rate it is common to introduce an initiator to the reaction system. The surfactant not only influences the nucleation of polymer particles, but the particle size distribution and the colloidal stability of the latex as well as the film properties are also affected by this component [2]. Another method commonly used to confer colloidal stability to a dispersed polymer is to place electrostatic charges at the particle surface; this is achieved by adding small amounts of functional monomers such as acrylic or methacrylic acid over the particle surface [3]. In addition, this monomer can promote the adhesion of the film to the substrate, which is of primary importance in polymer coatings [4].

In order to obtain high performance coatings, not only the nature of the polymer must be taken into account. The experimental conditions such as temperature, humidity, drying rate and nature of the substrate also play an important role in the film formation process [5]. Although latex films are widely used and extensively studied nowadays [6, 7], the exact mechanism involved in the transformation of a polymer dispersion into a coherent polymer film is not well established. Knowing this mechanism is important because it serves to design lattices so as to achieve latex films with desired properties.

Usually, the film formation process is arbitrary divided in three stages: 1) evaporation of water resulting in packing of latex particles, 2) their deformation, and 3) coalescence by interdiffusion of polymer chains between adjacent particles. The first stage is crucial because the structure achieved by the particles will remain in the film affecting its properties [8]. If the particles have enough time to pack, i.e. low evaporation rate of water, a dense packing could be obtained [9]. However, since the surface electrostatic charges of the polymer particles modify the interaction among them, this might affect the packing of the polymer particles.

The permeability and the water uptake of the latex film affect its performance because they can promote the film degradation and/or the damage of the substrate. When in contact with water, the coating tends to absorb water, swells and often the adhesion is lost or decreased [10]. Besides, small hydrophilic molecules, as surfactants employed in the polymerization reaction, pigments or water soluble oligomers, can be extracted from the film by water that increases the loss of film properties [11]. Also the correlation between latex film morphology, void content and film surface characteristics is relevant in paper coating performance [12].

In this work, the effect of varying the amount of AA on the latex film properties is studied from a theoretical point of view and compared to experimental data. A soft potential with one adjustable parameter is employed to model the change of the interparticle interaction due to the presence of AA and its effect on the film properties is investigated. The water vapor permeability as a function of time is studied experimentally and with the aid of a simulation approach. The surface of the simulated films is analyzed and correlated with their barrier properties. Finally, a comparison of experimental and theoretical water uptake, by means of film capacitance determinations, is presented.

## 2. Experimental

The monomers, n-butyl acrylate (BuA), methyl methacrylate (MMA), styrene (S) and acrylic acid (AA) technical grade were kindly donated by National Starch & Chemical. Ammonium persulfate (Fermont) and sodium bicarbonate (J.T. Baker) were used as initiator and buffer, respectively. Two surfactants, Abex 26 S ® (Rhodia) and DisponilALS 28 ® (Cognis) were used. Distilled water was used during the experiments and all reactants were used as received. The polymerization recipe is shown in Table 1. The amount of BuA and MMA was kept constant, however as AA was introduced in the formulation, the same quantity of S was removed, i.e. 98.4 g of S and 9.6 g of AA was used to prepare the copolymer with 4% of AA and so on. The amount of AA used in this study was 2, 4 and 6% wt of the monomer phase. The theoretical Tg calculated according to the Fox equation is close to 3 °C in all cases.

A semicontinuous emulsion polymerization reactor was used in the synthesis. The experimental device consisted in a 1 L glass reactor maintained at 80° C, under nitrogen atmosphere. In this reactor, the reagents corresponding to the main reactor load (shown in Table 1) were introduced, except the initiator solution. Once the reactor was maintained at the reaction temperature for 30 min, the initiator solution was added. After 10 min the addition of the pre-emulsion from the feeding tank started with the aid of a pump. The addition time was of 4 h in order to operate in starved-feed conditions and avoid secondary nucleation. After the addition time was over, the latex was maintained at 80° C for 1 h to reduce the residual monomer and left to cold to ambient temperature. All lattices were neutralized with a concentrated NaOH solution. The solid content in all cases was close to 40% wt as designed. A sample of the final dispersion of polymer particles was characterized by quasi-elastic light scattering with a LS Coulter 120 Nanosizer. The number average particle size was close to 400 nm in all cases with a polydispersity index less than 1.01

To obtain free polymer films, cleaned glasses of 10X10 cm$^2$ were covered with 3 mL of the dispersions and left to dry at laboratory conditions (21 °C and 50% relative humidity) for 5 days. Then, the polymer films were carefully peeled from the glass surface.

In order to measure the water vapor permeability of the films, the procedure described in ASTM E96 [13] was followed, specifically the wet cup test. In this method, a certain amount of water is placed in a glass container, which is sealed with the polymer film. The vials were maintained at the laboratory conditions and weighed as a function of time in a balance with 10 mg of error. Each sample was weighed three times and the mean value is reported. For each polymer film, three independent experiments were run.

Film capacitance was determined by means of Electrochemical Impedance Spectroscopy (EIS), which is a suitable technique to measure the film capacitance without removing the film from the substrate. Low carbon 6X8 cm$^2$ metal sheets were sanded with paper 600, cleaned and degreased with acetone; then 2.5 mL of the each dispersion was applied and left to dry under ambient conditions for 5 days, to give a film thickness around 160 μm with a deviation of 15%. In order to achieve a faster drying rate, the samples were located into a hermetic container with moisture adsorbent for 24 h, and then maintained at ambient conditions for 4 days. The impedance spectra were obtained with a Gill potentiostat, with amplitude of 10 ±mV between $10^4$ and $10^{-1}$ Hz, using graphite and saturated calomel as auxiliary and reference electrode, respectively. The spectra were obtained after 24 h of continuous immersion in a 0.5 M sodium sulfate solution. Data analysis was carried out by assuming a R(RC) equivalent circuit at high frequency range in a Zview software [14].

**3. Model description**
In our previous works [15, 16], a model of the first stage of the film formation process was developed. In those works, the polymer colloid particles were modeled as one-component fluid using the hard-spheres pair potential between the colloids. This simple potential is suitable for describing different colloidal features because it takes into account the excluded volume of the colloids [17, 18]; however, since colloidal particles have surface charge due to the presence of electrostatic charges, i.e. sulphate groups from the initiator and namely ionized carboxylic groups from the AA, colloidal particles start to repeal each other before they get into contact [19, 20]. As the number of surface charges on the particle surface increases, the repulsion between the particles becomes stronger. So, the augmentation of superficial charges due to the increment of the AA content may be modeled through an effective soft repulsive potential which also takes into account the compressibility of the species [21]. The soft potential between two particles is given by the following equation:

$$\beta U_{cc}(r) = \left(\frac{\sigma}{r}\right)^n \tag{1}$$

where $\beta=k_BT$, $k_B$ and T denote the Boltzmann constant and absolute temperature, respectively, $\sigma$ is the particle diameter considered as the unit length, $r$ is the center-to-center distance between particles and n is the softness parameter. The soft potential has a simple form but is able to reproduce the interparticle interaction in real systems. For example, in Fig. 1 a comparison between experimental and simulation results of the two-dimensional radial distribution function, g(r), is shown. The simulations results were obtained by calculating the g(r) with a canonical Monte Carlo (MC) simulation using potential (1) with n=16. The experimental structure was obtained by means of Confocal Scanning Laser Microscopy (CSLM) by Dullens *et al.*, taking the images of cross-linked poly(methyl methacrylate) particles suspended in TFH, in the first layer at the bottom glass wall of the sample container. [22]. These latex particles behave as soft spheres in THF [23]. As can be seen, there is an excellent agreement between experimental and simulation structure.

It is interesting to note that $\rho\sigma^2$, where $\rho$ is the two-dimensional number density, i.e. the number of particles divided by the area, is similar in experiment and simulation, 1.082 and 1.136, respectively. At larger distances the deviation between experimental and simulation g(r) results becomes more visible. The source of such difference could be the particle size polydispersity of the real system as well as the resolution of the microscope. [24]. The insert in Fig. 1 is a typical simulation configuration of the particles at equilibrium, which can be compared with the experimental picture. In both, a dense hexagonal packing is observed with low defect content. Based on the structure comparison we assume that soft interaction potential (1) with n around 16 is appropriate for modeling the effective interaction between charged latex particles.

As in previous works, the colloidal dispersion is modeled as one-component fluid, but this time using the soft pair potential, Eq. (1), instead of the hard-spheres potential between colloidal particles. The process of drying was simulated as follows. Initially $N$=4000 particles were introduced in a prism of $L_X=L_Y=25\sigma$ and initial length of $L_{Z0}=12.5\sigma$ (see Fig. 2a). The wall located on the left-hand side of Fig. 2a represents the hard impenetrable substrate. Each particle is tested to move *m* times according to Metropolis algorithm [25] (*m*=16 for the slow and *m*=8 for the high drying rate), with maximum particle displacement $l_p$=0.05 before the side $L_z$ reduces by $l_{zi}$=0.01$\sigma$ as showed by the arrows on the right-hand side of Fig. 2a. This mimics the evaporation of water and the consequent increment of the film density. The influence of the vapor-liquid interphase on the colloidal dispersion is modeled using a soft repulsive potential (Eq. (1), with $n_i$=8) that stands for the interaction between colloidal particles and the vapor-liquid interphase. Once a particle has been reached by the interphase (i.e. it acquires a particle-interphase repulsion larger than 100) it no longer experiments random motion because it is considered to be out the continuous phase. This trick reproduces the observed phenomena where a porous layer is formed at the vapor-liquid interphase as the drying proceeds. The simulation ends when all particles cannot move or $L_Z$=0. A detailed description of the model and the meaning of each parameter can be found elsewhere [15].

It is important to emphasize that the proposed model describes the first stage of the latex film formation, i.e. the particle packing during evaporation. However, at high densities the simulated particles may penetrate each other, which can be considered as a partial deformation and coalescence of the particles as the film formation goes on.

In order to evaluate the permeability of the simulated films, $N_a$=2000 small *a*-particles ("attacking particles") were placed right next to the film as illustrated in Fig 2b. The pair interaction between these "attacking particles" and dried film particles is given by Eq (1) with the following parameters $\sigma_{ac}$=0.5($\sigma_a + \sigma$) and $n_{ac}$=40, where $\sigma_a$ is the diameter of the *a*-particles ($\sigma_a$=0.01$\sigma$). There is no interaction between *a*-particles, they behave like ideal gas with respect to each other. Such approximation is valid for water vapor we study in the experiment. The *a*-particles were left to diffuse through the film moving preferentially to the substrate in a ratio 3:1. In each MC step all the *a*-particles were tried to displace once. This algorithm was run at least 10 times to calculate the average values. The number of MC steps needed for a given number of *a*-particles to reach the substrate will be related to the time that a given amount of water vapor needs to pass trough the experimental latex film.

The water uptake of the simulated films is determined by locating enough *a*-particles right next to the film to achieve certain density number, $\rho_a$=1. The interaction parameters between these particles and the ones that belong to the film are the same as mentioned above. Among the *a*-particles only the excluded volume is considered, i.e. they interact through the hard-spheres potential. This *a*-particle fluid was left to equilibrate and the adsorption isotherm, $\Gamma$, was calculated using the following equation,

$$\Gamma = \int_0^{z_{max}} \rho_a(z)dz \qquad (2)$$

where $\rho_a(z)$ represents the density profile of the *a*-particles and $z_{max}$ is film thickness, calculated from the dried film density profile.

Finally, film surface profiles were calculated according to a recently proposed methodology [26] which resembles an Atomic Force Microscope. That is, a "probe" ball of diameter 0.025$\sigma$ was moved perpendicularly to the substrate with a grid space of 0.05, in the *x* and *y* direction. Once the "probe" ball touches a particle of the film surface the coordinate is saved and employed to generate the film surface profiles and 3D images [27]. This information also can be employed to calculate the roughness of the latex films [28].

## 4. Results

The density profiles depicted in Fig. 3 show the effect of the softness parameter on the internal film structure. If the repulsion between particles is high, n=18, the obtained film is thicker if compared to other two films; as n decreases, the film thickness diminishes. In this analysis, it seems not to be a great difference using n=16.6 or n=16.8, however, the influence of this small variation is far from being negligible as will be seen below. It is interesting to note that the two outer layers, those near the vapor-liquid interface, are clearly different over all when n=18. For this case, the picks are smaller and broader than that obtained with lower values of n, which means that this film is not so well structured as the other two, and the difference in the properties may be related to this region.

The channels and defects in the film surface can be observed with microscopic techniques such as Scanning Electro Microscopy (SEM) and Atomic Force Microscopy (AFM). In order to obtain information about the surface of the simulated films an algorithm that mimics the operation of the AFM was developed [26]. In Fig. 4 a series of film surface profiles obtained from our simulations are showed. As we can see, the simulated film with the lowest value of the softness parameter, n=16.6, has a quite flat profile; it means that nearly all the particles lay in the same plane, increasing the surface density and diminishing the defects like channels and cracks. Such film may attain low roughness with a consequent high gloss [29]. On the other hand, as the repulsion among the colloidal particles become stronger (higher values of n) the roughness of the film increases with the diminution of the barrier properties[26]. The roughness is routinely measured with the AFM, and it is considered as the standard deviation from the mean value of the height. In Fig. 4, the mean is presented as a solid line and the standard deviation as a dotted line. In this figure it can be seen that film surface roughness diminishes as the softness parameter becomes lower.

In all cases, the layer directly over the substrate reaches the highest density but there is not a great visual variation of the structure achieved by this layer, as shown in the snapshots of Fig. 4. In the same figure, the coexistence of hexagonal and square arrays can be seen. The formation of these ordered arrays in the packing of particles is important since they provide dense regions with low presence of defects that do not allow the passage of water and aggressive species to the substrate and contribute to good mechanical properties of the film.

We have performed the comparison between experimental and simulation results of film permeability. Once the permeation curve (number of *a*-particles reaching the substrate as a function of MC steps) for n=18 was obtained, we have considered that 500 contacts with the substrate correspond to 0.67 g of water lost and 1670 MC steps correspond to 16 days of experimental time, defining a linear relation between the number of *a*-particles that reach the substrate and the water lost, as well as between MC steps and elapsed days.

The experimental and simulation results are given in Fig. 5. Experimental data revel that the lowest permeability is accomplished by the film with 2% wt of AA and n=16.6. This film has the lower values of permeability due to its compact structure and high density, as viewed from the density profiles. The next system, 4% wt AA and n=16.8 has a slightly higher permeability. The greatest value of permeability is achieved by the film with 6% AA and n=18. It is important to mention that the permeability simulation considers that all of the vapor transfer is carried out trough the voids left between the colloidal particles, not trough the polymer particles, since the hydrophobic core of the polymer particles does not contribute much to water transport as compared to voids and defects like channels and cracks of the film [30, 31]

The measurement of the water uptake using EIS techniques is based on the determination of the changes of the coating capacitance [32]. The capacitance is directly proportional to the dielectric constant, which for polymeric materials is close to 4-5, meanwhile, for water at 25° C it is close to 80. Therefore, any capacitance increment can be related to the presence of water within the film [33]. The measured capacitance of the latex films is showed in Fig. 6. As can be seen, increasing the AA content in the copolymer augments the amount of water up taken by the latex film. This increment could be a result of the morphology that the films achieve because of the AA addition, making stronger the interparticle repulsion and generating a less organized film structure or, due to the hydrophilic nature of the AA.

In order to determine which parameter has more influence on the film properties, the change of structure due to the addition of AA or the modification of the hydrophilic nature of the copolymer, the latex with 2% and 6% wt AA were dried at a higher drying rate by reducing the humidity of the system, which has a significant influence on the packing of particles. As seen in Fig. 6, at high drying rate the capacitance of the film formed with 2% wt AA substantially increases, reaching the same value as the film with 6% wt AA. More surprising is the fact that capacitance of the film formed with 6% wt AA does not depend on the drying rate. These results allow us to conclude that the packing of particles has strong effect on the water uptake by latex films, since water can penetrate the film mainly through its defects, pores or channels.

Fig. 7a shows density profiles of the *a*-particles adsorbed into the colloidal film. In the inner region of the film (close to the substrate), there is not any difference in the distribution of the *a*-particles; however, in the layers near the film surface one can observe that increasing softness parameter leads to a mayor presence of adsorbed *a*-particles due to the film structure in this region. The adsorption isotherms in Fig. 7b indicate the amount of *a*-particles that are inside the film, i.e. the integral of the density profiles of Fig. 7a, Eq. (2). We can observe that the value of $\Gamma$ increments as the softness parameter augments, which means that as the interparticle repulsion becomes stronger, there is an increment of the film water uptake. This algorithm also considers that adsorbed species are located at the interstices left by the polymer particles; therefore, the water uptake is predominantly influenced by the film morphology as a result of the interparticle interaction. In the same figure, it is shown that by increasing the drying rate the value of $\Gamma$ also increases and there is a minor influence of the softness parameter. Such trend is corroborated by our film capacitance measurements. Therefore, experimental and simulation results indicate that the influence of the interparticle repulsion on film properties is diminished as the drying rate increases, and the water uptake is mainly affected by the morphology of the film and not only by the addition of hydrophilic monomers.

## 5. Conclusions

In this work the modification of the film properties by the addition of acrylic acid is studied with experimental techniques and a simulation approach. It is shown that increasing the colloidal surface charge as a result of the addition of AA leads to latex films with a more porous structure which enhances the passage of water vapor. Also, it was demonstrated that films with higher surface roughness exhibit higher water vapor passage and water uptake. In addition, we found that increasing the amount of acrylic acid augments the film water uptake; however, we attribute such results to both the hydrophilic nature of the monomer and the latex film structure. Finally, it is important to mention that the theoretical approach presented in this contribution allows us to analyze the film surface characteristics, which is quite nontrivial to carry out in the case of soft latex particles [34].

**6. Acknowledgements**

Authors thank F. Vázquez (IMP) for helpful discussions. We also thank R.P.A. Dullens for kindly supplying his experimental data. Partial financial support from the Instituto Mexicano del Petróleo (project No. D.31519) is appreciated. Y.R. acknowledges his Ph.D. Scholarship from CONACyT-México.

**Table and Figure Captions**

Table 1. Composition of the polymerization formulation employed in the synthesis of emulsion polymers with different amount of acrylic acid.

Fig. 1. Comparison of the experimental (dots) and simulated (line) 2D radial distribution function, g(r). The softness parameter in the simulation is n=16. The insert is a snapshot of the simulation results which shows a dense hexagonal packing.

Fig. 2. a) Schematic representation of the simulation cell. The substrate is located on the left-hand wall. The liquid-vapor interphase of the colloidal dispersion moves as depicted by the arrows to simulate the water evaporation. b) Representation of the algorithm to determine the dried film permeability.

Fig. 3. Density profiles of totally dried films for different softness parameter.

Fig. 4. Two-dimensional profiles of the film surface for different values of n and snapshots of the layer directly over the substrate. The RMS (root mean square) of the surface height or film surface roughness is also reported.

Fig. 5. Experimental and simulation results of the film permeability as a function of time. Points correspond to experimental data and the solid lines are from simulation results.

Fig. 6. Film capacitance as a function of AA content.

Fig. 7. a) Density profiles of a-particles inside the colloidal film and b) adsorption isotherm as a function the softness parameter for different drying rates.

Table 1

| Substance | Main reactor (g) | Feeding tank (g) |
|---|---|---|
| n-butyl acrylate | 0 | 120 |
| Methyl methacrylate | 0 | 12 |
| Styrene | 0 | 103.2, 98.4 and 93.6 |
| Acrylic acid | 0 | 4.8, 9.6 and 14.4 |
| Ammonium persulfate solution 5% wt | 26 | 70 |
| Sodium bicarbonate solution 1% wt | 2 | 0 |
| Abex solution 10% wt | 10 | 50 |
| Disponil solution 10% wt | 10 | 50 |
| Water | 142 | 0 |

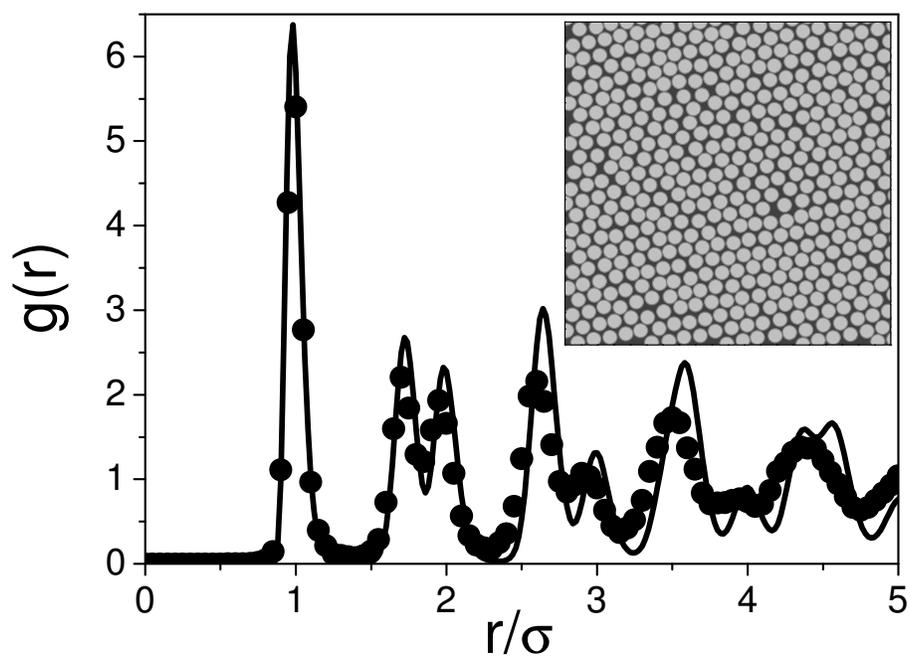

Fig. 1

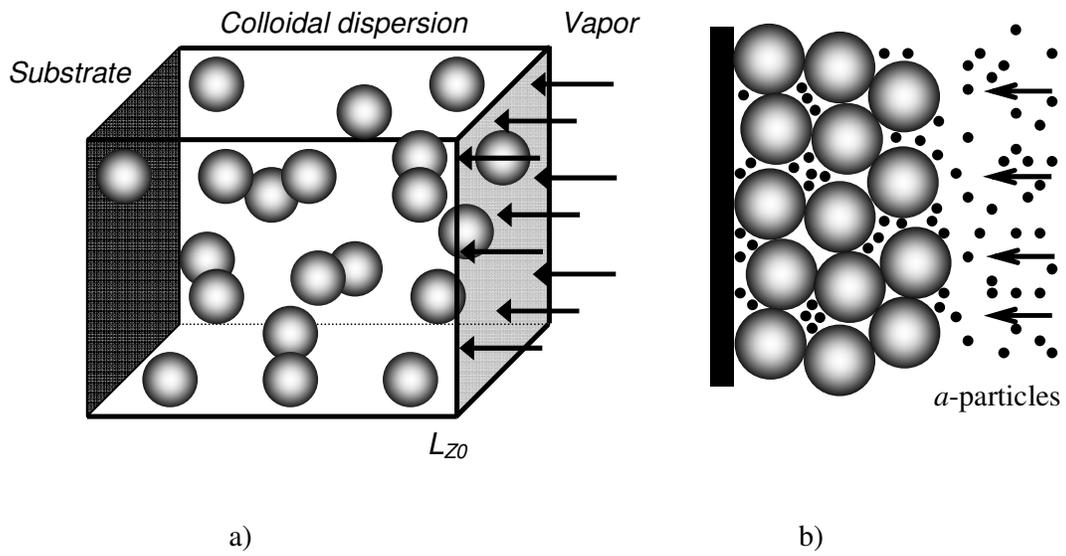

a) b)

Fig. 2

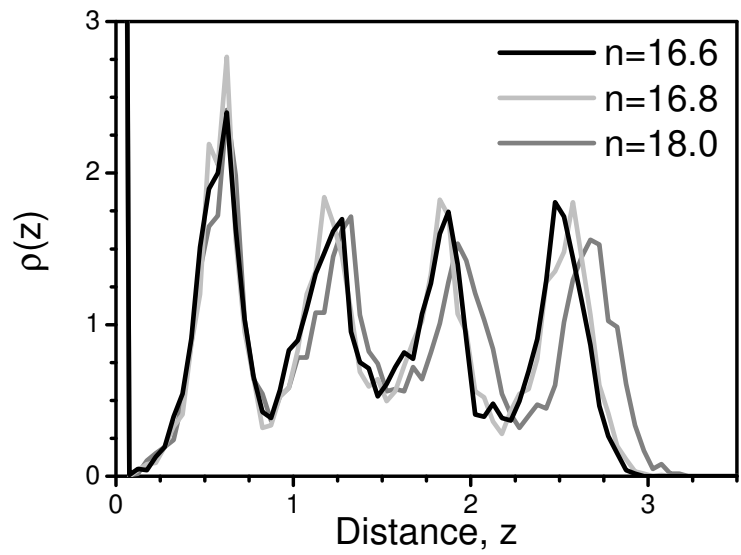

Fig. 3

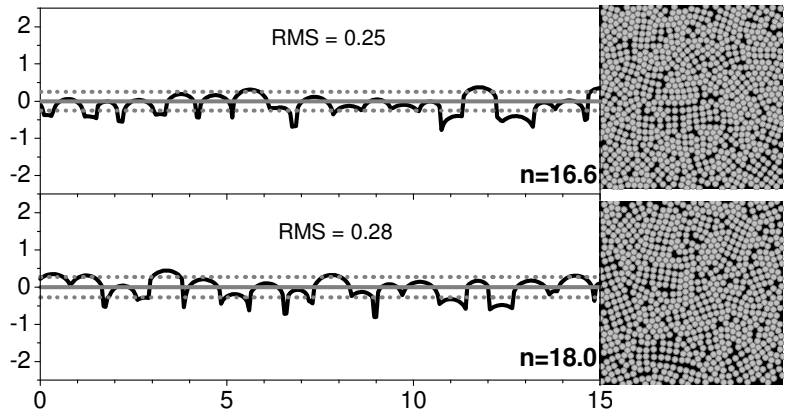

Fig. 4

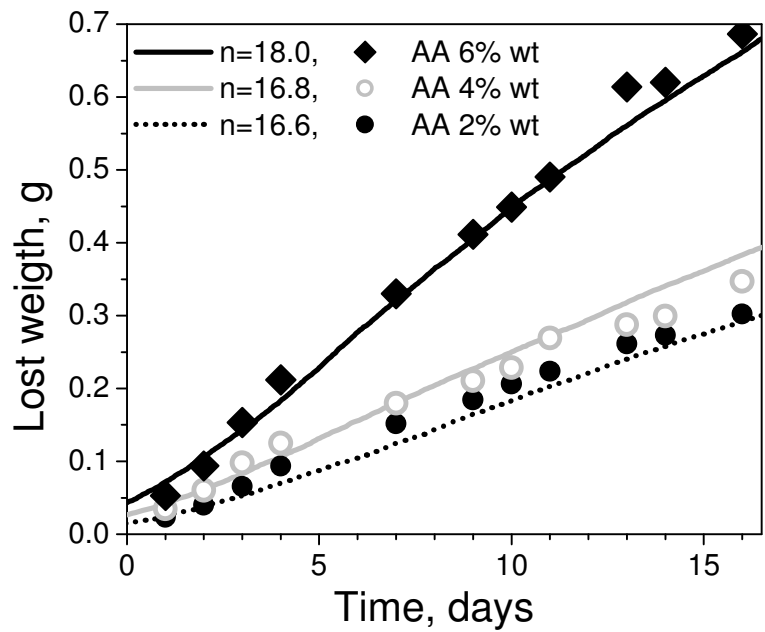

Fig. 5

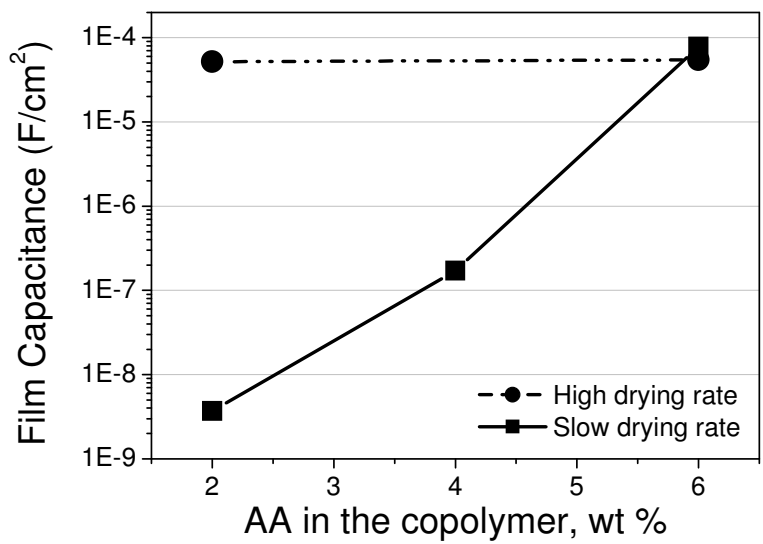

Fig. 6

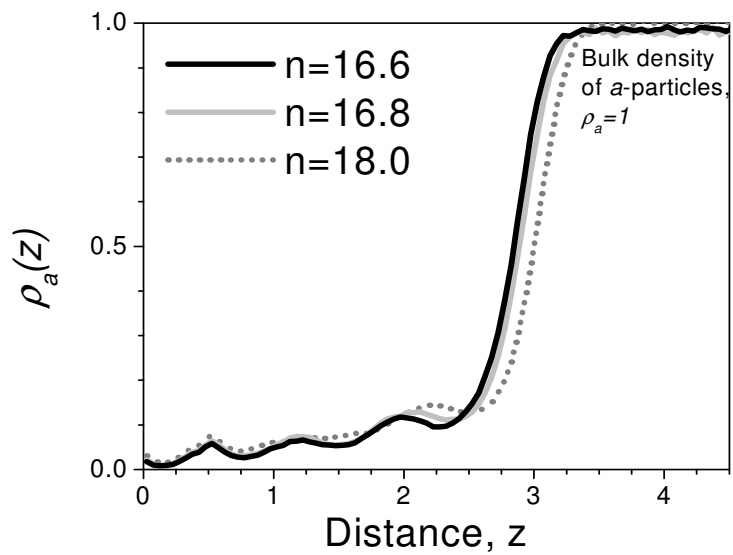

Fig. 7a

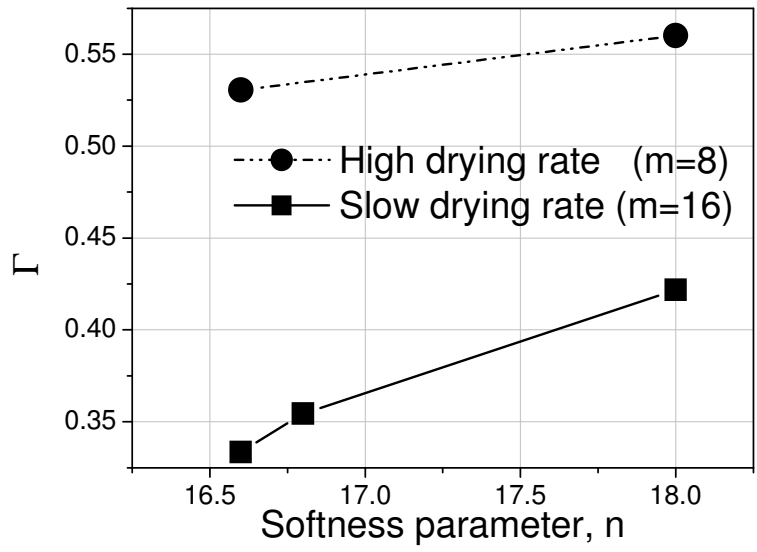

Fig. 7